\documentstyle[editedvolume]{crckapb}

\begin{opening}
\title{SETTING THE STAGE:\protect\\
       ULTRALUMINOUS GALAXIES IN A COSMOLOGICAL CONTEXT}

\author{TIMOTHY M. HECKMAN}
\institute{Department of Physics \&\ Astronomy\\
Johns Hopkins University\\
Baltimore, MD 21218 USA}

\end{opening}

\runningtitle{ULTRALUMINOUS GALAXIES \&\ COSMOLOGY}

\begin{document}

% The \begin{document} command comes after the \end{opening}
% command.

\begin{center}
{\bf Abstract}
\end{center}
I will try to put the ultraluminous galaxy phenomenon
into a broad cosmological context. Viewed from this perspective,
the significance of ultraluminous galaxies and the `starburst
vs. monster' debate becomes clear. Ultraluminous galaxies
are fascinating in their own right, allow detailed
study of the processes by which massive spheroids were built
and the IGM was heated and polluted, and resemble the most
luminous and dustiest galaxies at high-redshift. Ultraluminous galaxies
were apparently
far more common at z $\sim$ 3 than today. Recent
inventories in the local universe of the cumulative effect of nuclear burning
(metal production)
and of monster-feeding (compact dark objects in galactic nuclei)
imply that {\it either} stars or
monsters could have generated the observed far-IR cosmic background.
The starburst vs. monster debate has global, as well as local
importance.

\section{Introduction}
 
In this introductory talk I will address two questions. First, are
ultraluminous galaxies of fundamental significance, or just
interesting curiousities? Second, why do we care whether starbursts
or monsters dominate the energetics of ultraluminous galaxies?
I will argue that the answers to these questions are clear when addressed
in a global, cosmological context. In what follows, I will define
an ultraluminous galaxy to be a galaxy having a bolometric luminosity exceeding
10$^{12}$ L$_{\odot}$ for H$_0$ = 70 km s$^{-1}$ Mpc$^{-1}$
and a spectral energy distribution that is
dominated by rest-frame mid/far infrared emission.
 
\section{Are Ultraluminous Galaxies Important?}
 
\subsection{Motivation}
 
We now have a rather complete energetic census of the `local' (z $\leq$ 0.1)
universe. Comparing the luminosity functions of far-IR selected
galaxies, optically-selected galaxies, and quasars (cf. Soifer et al
1987) implies that ultraluminous IR-selected galaxies
are of comparable energetic significance to quasars of similar bolometric
luminosities, but are responsible for only of-order 1\% of the
far-IR emissivity (luminosity per unit co-moving volume element)
and $\sim$0.3\% of
the total bolometric
emissivity in the local universe. Thus, on simple energetic grounds
it might indeed be possible to dismiss 
ultraluminous galaxies as merely intruiging
oddities.
 
The generic response to such a dismissal is to reply that
although they are rare, ultraluminous galaxies are
excellent local laboratories in which the physics and phenomenology
of galaxy building can be studied in far greater detail than at
high redshift. Some of the lessons that we have learned from the
investigation of 
local ultraluminous galaxies have potentially wide-ranging implications.
These include the role played by galactic mergers in making (some/all?)
elliptical galaxies (cf. Schweizer 1997),
the apparent efficacy with which such mergers
can transport much of the ISM of the merging galaxies into the
circumnuclear region (e.g. Mihos \& Hernquist 1994), the subsequent
triggering of circumnuclear
star-formation at a rate approaching the maximum allowed by
physical causality ({\it viz.} SFR $\sim$ M$_{gas}$/t$_{cross}$ - Heckman
1993),
and the resulting heating and metal-enrichment of the inter-galactic medium
by galactic `superwinds' that are driven by the
collective
effect of the millions of supernovae and stellar winds in the starburst
(cf. Heckman et al 1996).
 
On the other hand, given the current paradigm of the
hierarchical assembly
of galaxies in which (to quote Simon White) `Galaxy formation is
a process rather than an event', it is a fair and certainly
germane question to ask whether {\it typical} galaxies ever go
through an ultraluminous phase. More precisely: could the ultraluminous
phenomenon be an integral part of the formation of galactic spheroids
(ellipticals and bulges)? Is Arp 220 really a galactic 
`Rosetta Stone' or just an `(infra)red herring'?
 
The recent stunning advances in the observation
of galaxies at high redshift mean that we can finally start to
answer these questions directly.
 
\subsection{The Lyman Break Galaxies as Ultraluminous Galaxies}
 
Dickinson (1998) has recently published an ultraviolet
(1500 \AA\ rest-frame) luminosity
function for a large sample of galaxies at z $\sim$ 3 from
the Hubble Deep Field and larger but shallower ground-based surveys,
selected on the basis of their rest-frame UV spectral energy distributions
(the `U drop-out' or `Lyman Break' galaxies). Ignoring any correction
for dust extinction, this luminosity function would imply that galaxies
with apparent UV luminosities exceeding 10$^{12}$ L$_{\odot}$ are
exceedingly rare in the early universe, with co-moving space densities
of-order 10$^{-4}$ that of present-day 
normal Schechter L$_{*}$ galaxies.
 
Far be it from me to propose at a conference laden with infrared
astronomers that we actually ought to ignore the effects of dust on these
results! Indeed, the generation of techniques to correct the Lyman Break
galaxies for the effects of extinction has evolved into a virtual
cottage industry (e.g. Madau, Pozzetti, \& Dickinson 1998;
Meurer, Heckman, \& Calzetti 1999; Pettini et al 1998; Sawicki \&
Yee 1998). Daniela Calzetti will give a report from the front lines
on this issue later in this conference. To summarize, plausible
values for the mean/typical UV extinction suffered by the Lyman Break
galaxies range from 1 to 4 magnitudes.
The UV color-
magnitude relation for the Lyman Break galaxies in which the
fainter galaxies are bluer (Dickinson, private communication)
is reminscent of the strong dependence of extinction on luminosity
seen in local starbursts (Heckman et al 1998). This implies that
a correction of the observed UV luminosity function of the Lyman
Break galaxies for extinction
will change the shape of the function, and not merely its
normalization in luminosity. This is clearly seen at low-redshift
(Buat \& Burgarella 1998).
 
Meurer, Heckman, \& Calzetti (1999)
have made a rough attempt to correct the Lyman Break luminosity
function at z $\sim$ 3 for the effects of luminosity-dependent extinction.
Their results imply that galaxies with {\it intrinsic} UV
luminosities of 10$^{12}$ L$_{\odot}$ are actually rather
common at z $\sim$ 3, with a co-moving space density that is
of-order 10$^{-1}$ that of present-day Schechter L$_{*}$ galaxies. The
luminosity/extinction correlation means that the
spectral energy distributions of the most luminous Lyman
Break galaxies should then be dominated by the infrared, so they
would meet my definition of `ultraluminous galaxies'. Thus,
the co-moving space density of ultraluminous galaxies at
z $\sim$3 would be similar to the space density of M82-level
starbursts today.
 
\subsection{The SCUBA Sources in Context}
 
ISO and especially SCUBA have opened a new window on the early universe
and allowed us to make the first direct comparisons of the far-IR properties
of the universe of today to the distant past. The presentation
of these marvelous new results will constitute a major portion of this
conference, so I will keep my remarks brief.
 
While the distribution of the SCUBA sources in redshift is still a matter
of on-going investigation (e.g. Lilly et al 1998; Trentham, Blain, \& Goldader
1998; Smail et al 1998),
it is clear that they constitute a major
new population of objects that are energetically significant in
a cosmological context.
As described above, Meurer, Heckman, \& Calzetti (1999) have used
empirical methods for
correcting the Lyman-Break population at z$\sim$ 3 for extinction. The
extinction-corrected intrinsic UV luminosity function they derive implies that
the most luminous Lyman-Break galaxies may overlap the SCUBA sub-mm population
in luminosity and space-density.
 
Heckman et al (1998) have shown that local starbursts obey quite strong
relations between such fundamental parameters as luminosity, metallicity,
extinction, and the mass of the galaxy hosting the starburst.
More massive galaxies host more metal-rich
starbursts, which are in turn more heavily extincted by dust.
This probably reflects the well-known mass-metallicity relation for galaxies
and the roughly linear dependence of the dust/gas ratio on
metallicity. Moreover - and as noted above - the more
luminous local starbursts are more heavily extincted by dust, and the
UV color-magnitude relation for the Lyman Break galaxies suggests this may
also be true at high-redshift.
 
It therefore seems plausible that the SCUBA sources at high-z
are the high-luminosity tail of the Lyman-Break population,
and probably represent the most metal-rich (dustiest) starbursts
occuring in the most massive halos. This idea is certainly
consistent with a strong similarity between the high-z SCUBA sources
and local ultraluminous galaxies.
 
\section{Starbursts Versus Monsters: A Global Inventory}
 
Of course, the over-arching theme of this conference is the
debate over the nature of the fundamental energy source
in ultraluminous galaxies: starburst or monster? In the spirit
of the rest of my talk, I'd like to consider the issue
of the relative energetic significance of stars vs. monsters
from a global perspective. Andy Lawrence developes many of the
same themes in his contribution to this conference.
 
First, we can conduct an inventory of the luminous energy present in the
universe today. This represents the cumulative effect of the production of
luminous energy over the history of the universe (primarily by
stellar nuclear-burning and accretion onto supermassive
black holes), diminished only by the $(1+z)$ stretching of the photons.
This inventory is made possible by the recent ultra-deep
near-UV-through-near-IR galaxy counts in the Hubble Deep Field
(Pozzetti et al 1998) on the one hand, and the landmark detection by COBE of
a far-IR/sub-mm cosmic background on the other (Puget et al 1996;
Hauser et al
1998; Schlegel, Finkbeiner, \& Davis 1998).
 
The total present-day energy density contained in the cosmic IR
background
is $\sim$ 6 $\times$ 10$^{-15}$ erg cm$^{-3}$ (Fixsen et al 1998), which is
comparable
to the total energy density contained in the NUV-through-NIR light
due to faint galaxies (Pozzetti et al 1998). The origin of the latter
is clear: the light of these faint galaxies
is overwhelmingly due to ordinary stars (nuclear fusion).
However, the origin of the cosmic IR background is {\it not}
so clear. As I will outline below, simple `from-first-principles'
arguments imply that this luminous energy may have been generated
predominantly by either stars or monsters that were deeply
shrouded in dust.

One obvious way to evaluate whether stellar nucleosynthesis could
have been responsible for producing the energy contained in
the cosmic IR background is to take an inventory of the byproducts
of nuclear burning in the local universe. The recent compilation 
assembled by Fukugita, Hogan, \& Peebles (1998) implies that the baryonic
content of galaxies, the intracluster medium, and the general inter-galactic
medium is $\Omega_B$ $\sim$ 4.3 $\times$ 10$^{-3}$, 
2.6 $\times$ 10$^{-3}$, and 1.4 $\times$ 10$^{-2}$ respectively.
If we adopt a mean metallicity of 1.0, 0.4, and 0.0 Z$_{\odot}$
for these respective baryonic repositories and use the estimate
due to Madau et al (1996) that each gram of metals produced corresponds to
the generation of 2.2 $\times$ 10$^{19}$ ergs
of luminous energy, the implied co-moving density of
energy produced by nuclear burning
is then 2 $\times$ 10$^{-14}$ erg cm$^{-3}$. If we instead
assume that the ratio of metals inside galaxies to those outside galaxies
is the same everywhere as it is clusters of galaxies
(cf. Renzini 1997), then the
total mass of metals today is about twice as large as the above estimate,
as is the associated luminous energy. To compare
these values to the cosmic IR background, we need to know the mean
energy-weighted redshift at which the photons in the IR background
originated. Taking $<z>$ = 1.5, the resulting observable energy density in the
present universe would be in the range
8 to 16 $\times$ 10$^{-15}$ erg cm$^{-3}$. This is comparable to the
sum of the energy contained in the IR plus the NUV-through-NIR
backgrounds. Thus, there is no fundamental energetics problem with
a stellar origin for the cosmic IR background.

What about dusty quasars? 
At first sight, this
does not appear to be a plausible source for the bulk of the
cosmic IR background. The cumulative emission from the known
population of quasars - selected
by optical, radio, or X-ray techniques - has resulted in a bolometric
energy density today of about 3$\times$ 10$^{-16}$ erg cm$^{-3}$
(cf. Chokshi \& Turner 1992), only about 5\% of the cosmic
IR background. But, what if there exists a substantial population
of objects at high-redshift that are powered by accretion onto
supermassive black holes, but which are so thoroughly buried in dust
that they radiate primarily in the IR, and have thus far been missed
in quasar surveys? That is, could the cosmic IR background be
produced by a population of monster-powered ultraluminous galaxies
at high redshift? Might the SCUBA sources be our first glimpse of
this population? Could this same population of
dust-enshrouded AGN be responsible for the bulk of the cosmic
hard X-ray background (as Fabian et al 1998 have argued)?
 
One way to assess whether accretion onto supermassive black holes
is an energetically feasible source for the observed cosmic IR
background is to examine the fossil record in nearby galaxies.
The generation of the cosmic IR background by the accretion
of matter onto supermassive black holes necessarily implies
that the centers of galaxies today will contain the direct evidence
for this accretion. Is there enough mass in the form of supermassive
black holes in galaxies today to have produced the IR background?
 
Recent dynamical surveys of the nuclei of nearby galaxies strongly
suggest that supermassive black holes are common or even
ubiquitous, with a mass that is $\sim$ 0.5\% of the stellar mass
of the spheroid (bulge or elliptical) within which the black hole
resides (Magorrian et al 1998; Richstone et al 1998). The
corresponding ratio of black hole mass to spheroid blue luminosity
in solar units is roughly 0.045 for a Schecter L$_{*}$ elliptical.
Fukugita, Hogan, \& Peebles (1998) estimate that the present-day blue
luminosity density associated with spheroids is 4.6 $\times$ 10$^7$
L$_{\odot}$ Mpc$^{-3}$, so the implied mean density in the
form of supermassive black holes is $\sim$ 2 $\times$ 10$^6$
M$_{\odot}$ Mpc$^{-3}$.
 
If we assume that accretion onto a supermassive black hole
releases luminous energy with an efficiency $\epsilon$ = 10\%
($E = \epsilon M c^2$),
the present-day black hole mass density implies a total production
of 1.2 $\times$ 10$^{-14}$ erg cm$^{-3}$ in co-moving coordinates.
If the energy-weighted mean redshift at which this was emitted is
$z \sim$ 2, the present-day luminous energy density is then
4$\times$ 10$^{-15}$ erg cm$^{-3}$. This is roughly an order-of-magnitude
larger than the luminous energy produced by the known quasar population,
but matches the energy contained in the cosmic IR background rather well.
 
There are therefore three possible interpretations of this. First,
we may have substantially over-estimated the mass of black holes in
the nuclei of galaxies today. A recent analysis by van der Marel
(1999) yields an mean ratio of black-hole-mass to spheroid luminosity
that is a factor of 2 to 3 smaller than the Magorrian et al value.
Second, the formation of a supermassive
black hole may occur with a mean efficiency for the production of
radiant energy that is small (e.g. 1\% rather than 10\%). Perhaps
the quasar phase corresponds to high efficiency and produces most of the
radiant energy, but most of the accretion and black hole growth 
produces very little radiation (e.g. Narayan 1997).
Third, maybe the cosmic IR
background does have a substantial contribution from dust-enshrouded
`monsters'. If true, this would imply that over the history of
the universe, monsters have produced as much luminous energy
as stars!
 
\section{Summary}

When examined from a global, cosmological perspective, the answers
to the two questions I posed in the Introduction seem clear:

{\bf 1. Are ultraluminous galaxies of fundamental significance, or just
interesting curiousities?}

Ultraluminous galaxies are spectacular and fascinating in their own
right. They are unique local laboratories that allow the detailed
investigation of the physical processes by which galaxies were
built and by which the intergalactic medium was heated and
chemically-enriched. The most luminous members of the Lyman Break
galaxy population at high-redshift are almost certainly ultraluminous systems
dominated by far-IR emission and the SCUBA sources at high-z
(probably the most metal-rich, dustiest starbursts
occuring in the most massive halos) resemble local
ultraluminous galaxies. Thus, dusty ultraluminous galaxies have
been responsible for a significant fraction of the high-mass star-
formation and associated metal production at early times.

{\bf 2. Why do we care whether starbursts
or monsters dominate the energetics of ultraluminous galaxies?}

We now know that the cosmic IR background contains as much energy
as the integrated UV, visible, and NIR light from all the galaxies
in the universe. Recent inventories of the by-products of both nuclear
burning (metals and post-big-bang He) and of black hole accretion
(dark compact objects in galactic nuclei) in the present-day universe
imply that {\it either} a population of dusty star-forming galaxies
{\it or} of dust-enshrouded monsters could have readily produced the IR
background. Thus, on a global scale, the `starburst vs. monster'
debate is of central importance. It is possible that - integrated
over cosmic time - accretion onto supermassive black holes has
produced as much total radiant energy as nuclear burning in stars.
Future multi-wavelength observations of the sources detected by
ISO, SCUBA, and SIRTF will go a long ways towards settling this 
crucial issue.\\[5mm]

{\bf Acknowledgments}

I would like to thank the Local Organizing Committee (particularly
Reinhardt Genzel, Dieter Lutz, and Linda Tacconi) and the staff
of the Ringberg Castle for making this meeting stimulating,
enjoyable, hastle-free, and so civilized! I also congratulate
Bob Joseph and Dave Sanders for their spirited debate (although
they're both lousy soccer players).
This work was supported in part by NASA LTSA grant
NAGW-3138.

\end{document}